# Class Based Admission Control by Complete Partitioning -Video on Demand Server

Soumen Kanrar[1] and Mohammad Siraj[2]

[1]Department of Computer Engineering ,College of Computer and Information Sciences, King Saud University, Riyadh -11543, SA
`Soumen_kanrar@yahoo.co.in`

[2]Department of Computer Engineering ,College of Computer and Information Sciences, King Saud University, Riyadh -11543, SA
`siraj@ksu.edu.sa`

## ABSTRACT

*In the next generation network (NGN) environment specific consideration is on bandwidth minimization, because this reduces the cost of network. In response to the growing market demand for multimedia traffic transmission, NGN concept has been produced. The next generation network provides multimedia services over high speed networks, which supports DVD quality video on demand. Although it has numerous advantages, more exploration of the large-scale deployment video on demand is still needed.*

*The focus of the research presented in this paper is a class based admission control by the complete partitioning of the video on demand server. In this paper we present analytically and by simulation how the blockage probability of the server significantly affects the on demand video request and the service. We also present how the blockage probability affects the performance of the video on demand server.*

## KEYWORDS

Next Generation Network, Video on demand server, Blockage Probability, partition.

## 1 Introduction

Video-on-Demand (VoD) systems are expected to be one of the most important services supported by the next generation of high speed networks, video servers, and distributed multimedia file system. Typically a large number of video files are stored in a set of centralized video servers and played through high-speed communication networks . The geographically distributed clients will be able to submit their request for a video from any place at any time through the network. Due to stringent response time requirements , continuous delivery of a video stream has to be guaranteed by reserving an I/O stream and an isochronous channel needed for the delivery. In this paper we consider the complete partition of the server for efficiently handling the client request. In this work we present the system performance of video on demand server with respect to the blockage probability at the individual VoD servers. Over the last two decades researchers have focused on analyzing the packet loss, packets delay, jitter to find the performance of the different types of network system[ 1 ]. To maximize the utilization of these channels, efficient scheduling techniques have been proposed by Vin and Rangan [18],Ozden et al. [19, 20], Freedman and Dewitt[21], Keeton and Katz [22], Oyang et al. [23] just to name a few. These techniques are sometimes referred to as user centered [24, 25] in the sense that the channels are allocated among the users. Although this approach simplifies the implementation, dedicating a stream for each viewer will quickly exhaust the network – I/O bandwidth at the server communication ports. In fact the Network –I/O bottleneck has been observed in many systems, such as Time Warner cable's Full Service Network project in Orlando , Microsoft's Tiger video Fileserver [4-26] and so.

Previous research in VoD systems have focused on the disk scheduling, disk stripping, video block placement, admission control at the level of disks and disk groups [2], [3], [4], [5]. Chen et al [6] introduces the concept of distinguishing between high and low priority classes but the resource capacity is partitioned between the two classes. Admission control is enforced on the dedicated partition of server capacity as a separate M/M/C/C queue. Therefore, a cost effective design for the VoD system, needs to evaluate a collection of various VoD system components [7]. In terms of VoD transmission network, the system design must guarantee the required bandwidth for video traffic.This bandwidth must support the VoD service quality needs to meet its packet loss polices. Video content delivery consumes large amounts of bandwidth in the networks, due to its scalability. In addition, system client must comply with the necessary buffer size and video request for the VoD delivery policy [8]. O vector et al. [9] develops a performance evaluation tool for the system design and a user activity model to describe the utilization of the network bandwidth and video server usage. It becomes important to get better performance from the system by allocating the dedicating channel to the user  through the fixed ports. With the increase of the population size the load on the server becomes heavy. It is essential to divide the ports into smaller groups  and allocated the individual smaller group to particular class of user.

This paper is structured as follows: Section 2 introduces the Network architecture of video on demand system which briefly represents the incoming traffic pattern to the video on demand server . Section 3 illustrate the  analytic from of the 'Admission control' based  on the partition of the ports of  video server. Section 4 present the parameters description of the  simulation environment. Section 5 presents the simulation result with respect to the proposed algorithms .

Section 6 and 7 presents conclusion remarks and acknowledgements.

## 2.Network Architecture (VoD system )

The request comes from the client to the VoD server for the two types of movies one for the popular movies and other for the unpopular movies. The request also is for two types in the VoD Network. The first one request is for initializing or starting the video movie. The other type is the request for interactive service (e.g. stop/pause, jump forward, fast reverse etc) to be performed on the viewed movies. Since each of these request is independent from each other, and arrival requests come from large numbers of client set-up terminals, the arrival process of normal requests as well as of interactive requests to the video server can be modeled as a Poisson distribution with average rates $\lambda_S$ ( for steady session) and $\lambda_I$ ( for interactive session) respectively. With this assumption, the distribution of the sum of $K$ of independent identically distributed random variables, representing the request inter – arrival times (Which are exponential distributed mutually independent random variables ) is then the Erlang distribution.
 A typical VoD architecture consists of three critical subsystems : single or cluster video servers, high speed wide-area and local distributed networks and many users populations. The hierarchical VoD system architecture used in this analysis consist of local and remote sites. Each site is characterized by a cluster of video servers. The video servers delivers high quality digitized multimedia data to clients over local distributed networks from local sites or over high speed networks from remote sites. The set-top boxes at the client site provide the decoding and display functionality, in addition to providing buffer for periodically delivered video segments from the video servers. The organization of local and remote clusters in reference to the client (user) population is shown in figure-1

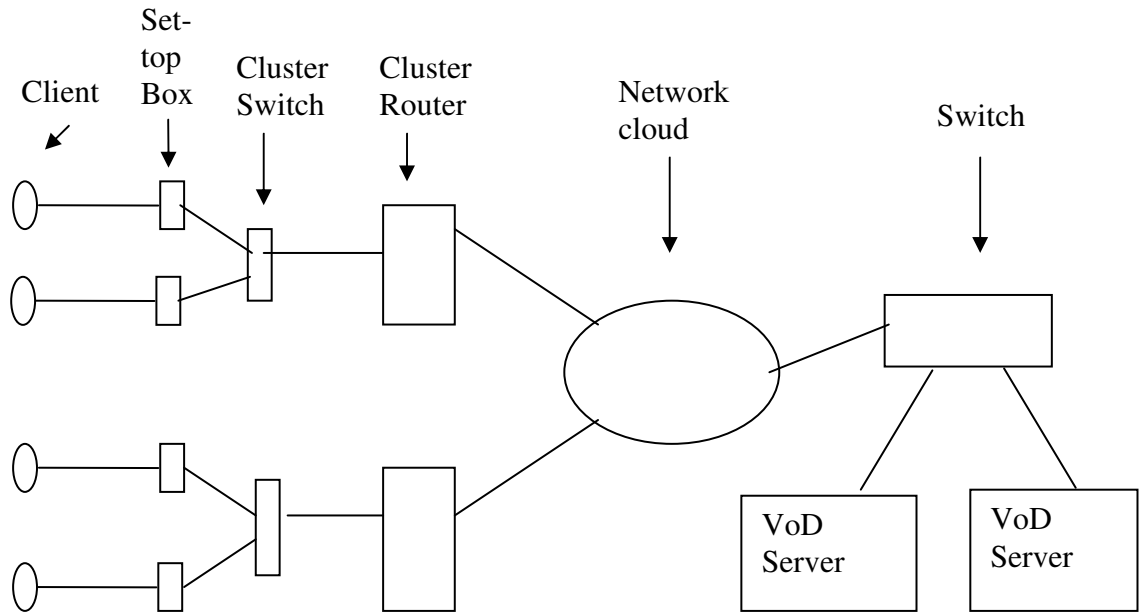

Figure – 1: Video-on-demand Network

Each local cluster is dedicated to its reference user population. The local cluster may store a complete or a partial set of videos from the video collection. The service from the local cluster is provided over a local distribution network, such as an ATM LAN. The requests originating from a reference user population are best served by the local site because of the absence of network contention. The clusters of servers have limited bandwidth and limited ports and therefore reject request when there are no free port.

The distribution of VoD movies request generally follows a Zipf-like distribution, where the relative probability of a request for $i$ (the most popular page) is proportional to $1/i^\alpha$, with $0 < \alpha < 1$ and typically taking on some value less than unity. The request distribution rarely follows the strict Zipf law (for which $\alpha = 1$) [10]

For Zipf-like distribution, the cumulative probability that one of the M popular movies is accessed (i.e. the probability of a popular movies request) is given asymptotically by:

$$\Psi(M) = \sum_{i=1}^{M} \frac{\delta}{i^\alpha} \approx \delta \frac{M^{1-\alpha}}{(1-\alpha)} \quad \ldots\ldots(1)$$

and $\delta \approx \frac{(1-\alpha)}{N^{(1-\alpha)}}$ Where N is the total number of movies in the system. $\Psi(M)$ can be

approximated as: $\Psi(M) \approx (M/N)^{(1-\alpha)}$ …………(2)

because $(M/N) < 1$ for all meaningful $M$ and a large $\alpha$ increase $\Psi(M)$. Most requests are concentrated on a few popular movies. Based on this we can estimate the probability of a request for an unpopular movie for a VoD system with $N$ total movies and $M$ popular one as

$p_{un} = 1 - (M/N)^{(1-\alpha)}$ ……….(3)

## 3. Admission control based on Complete (partition) video server

A capacity based admission control test at the server determines if the arriving request can be admitted into the system. The server is able to serve a maximum of n requests in a round and $n_k$ is the number of requests currently in service.

In this section, we develop an analytical model for evaluating the performance of a single video server under complete (partition) sharing of the server.

Let C is the total numbers of port in the server and the ports are divided into K class of partition such that

$$C = \sum_{j=1}^{k} c_j \quad ----- (4)$$

and the request coming from the class of population with rates $\lambda_i$ where $i \in ¥$

**Case -1:**

Here $C_1$ is the total numbers of ports for the 1$^{st}$ partition. $Q_i^{(1)}$ is the total ports currently occupied ( or busy) for the 1$^{st}$ partition of the server. The probability of the event that $i^{th}$ incoming request coming from the 1$^{st}$ population- class get free port to the 1$^{st}$ partition of the server is

$$\frac{1}{k} \cdot \frac{C_1 - Q_i^{(1)}}{C_1} \cdot ---------- (5)$$

Now if the 1$^{st}$ partition of the server is blocked i.e. there is no free port when, $C_1 = Q_i^{(1)}$

Let $B_1$ is the event that the 1$^{st}$ partition is blocked. Then probability of the event ($B_1$, 1$^{st}$ partition of the server is blocked)

$$p_b(X = B_1) = \frac{\frac{E_1^{C_1}}{(c_1)!}}{\sum_{j=0}^{c_1} \frac{E_1^j}{(j)!}} \quad ---------------- (6)$$

Total traffic load for the 1$^{st}$ partition (according to little theorem) is $\sum_{i=1}^{C_1} \lambda_i h_i$ and the corresponding Erlang is $E_1 = \frac{1}{T_1}\left[\sum_{i=1}^{C_1} \lambda_i h_i\right]$ where $h_i$ is the required time occupied the port by the $i^{th}$ customer and $T_1$ is the total time that the every ports of the 1$^{st}$ partition is busy.

**Case 2**

When every ports of $C_1$ Partition is blocked and the request forward to the next block.

Let $B_1$ is the event that the partition $C_1$ is blocked and A is the event that $C_2$ .i.e. 2$^{nd}$ partition has free port.

$$p(A/B_1) = \frac{p(AB_1)}{p(B_1)} = \frac{p(A)p(B_1)}{p(B_1)} = p(A) \quad ------- (7)$$

since $(A \cap B_1) = \Phi$ and A, $B_1$ are mutually exclusive independent events. $p_b(B_1)$ is the blocking probability for the $C_1$ partition and $p(AB_1)$ represent the probability of the event ( $C_1$ partition blocked but $C_2$ partition have (at least one) free port). i.e. $C_2 f Q_i^{(2)}$

$$p(A/B_1) = (1-\frac{1}{k}).\frac{1}{k}.\frac{C_2 - Q_i^{(2)}}{C_2} \quad \text{----------- (8)}$$

Total traffic load for the 2$^{nd}$ partition is $\sum_{i=1}^{C_2} \lambda_i h_i$ and the Earlang of that partition can be represented as

$$E_2 = \frac{1}{T_2}\left[\sum_{i=1}^{C_2} \lambda_i h_i\right],$$ where $h_i$ is the time occupied by the ports by the $i^{th}$ customer from the population. $T_2$ is the total time that the every ports of the 2$^{nd}$ partition busy.

Now the blocking probability for the 2$^{nd}$ partition is

$$p_b(X = B_2) = \frac{\frac{E_2^{C_2}}{(C_2)!}}{\sum_{j=0}^{C_2}\frac{E_2^j}{(j)!}} \quad \text{------------ (9)}$$

**General case**

If request comes from $i^{th}$ class of population and served by the $j^{th}$ partition block of the server clearly $1 < j$

Now,

$$p(A_j / B_i B_{i+1}..............B_{j-1}) = \frac{p(A_j B_i B_{i+1}..........B_{j-1})}{p(B_i B_{i+1}........B_{j-1})} = \frac{p(A_j)p(B_i)p(B_{i+1}).......p(B_{j-1})}{p(B_i)..............p(B_{j-1})} = p(A_j) =$$

$$(1-\frac{1}{k})^{j-1}.\frac{1}{k}(\frac{C_j - Q_i^{(j)}}{C_j})\ldots\ldots\ldots(10)$$

Where $A_j B_i B_{i+1}............B_{j-1}$ is the event that the previous all partition from $i$ to $j-1$ is blocked only $j^{th}$ partition has at least one free port.

$B_i B_{i+1}.......B_{j-1}$ represent the event that all the partition from $i$ to $j-1$ is blocked.

$$p_b(B_i B_{i+1}.......B_{j-1}) = p_b(B_i)p_b(B_{i+1}).........p_b(B_{j-1}) \ldots\ldots\ldots\ldots(11)$$

because $p_b(B_m \cap B_n) = \Phi$, $\forall (m,n) \in N$ and $m \neq n$ where $i \leq m \leq j-1, i \leq n \leq j-1$.

Expression (11) can be represent as

$$p_b(B_i B_{i+1}.......B_{j-1}) = \left[\frac{\frac{E_i^{C_i}}{(c_i)!}}{\sum_{k=0}^{c_i}\frac{E_i^k}{(k)!}}\right]\left[\frac{\frac{E_{i+1}^{C_{i+1}}}{(c_{i+1})!}}{\sum_{k=0}^{c_{i+1}}\frac{E_{i+1}^k}{(k)!}}\right]................\left[\frac{\frac{E_{j-1}^{C_{j-1}}}{(c_{j-1})!}}{\sum_{k=0}^{c_{j-1}}\frac{E_{j-1}^k}{(k)!}}\right]\ldots\ldots\ldots\ldots(12)$$

and, $E_n = \frac{1}{T_n}\left[\sum_{i=1}^{C_n} \lambda_i h_i\right]\ldots\ldots\ldots\ldots(10)$

Here we present the algorithm for request forwarding to the free port of the video on demand server.

```
// V be the set of all incoming request comes from the user
// population with different incoming rate, $V = \bigcup_{1}^{n} V_i$
// Let $\lambda_i$, $1 \leq i \leq l$ be the incoming traffic request from $V_1$
// sub space such that cardinality of $(V_1) = l$
// $C = \sum_{j=1}^{k} C_j$, $C_1$ be the number of ports belongs to the 1st
// partition of the VoD server. $k$ stand for number of partition
// $C$ is the total number of ports of the VoD server. $Q_i^{(j)}$
// represents the number of free ports of the $j^{th}$ partition of the
// server for $i^{th}$ request from $V$ population.
// $C_i \in A[1..k]$ in 1D space
// $Q_j^i \in B[1..k, 1..k]$ in 2D space
    Var
      i, j =1 : integer
      for i =1 to k do begin
        if $C_1 > Q_i^{(1)}$
        // $i^{th}$ request be served
        then    $Q_i^{(1)} = Q_i^{(1)} + 1$
        else
        // request forwarded to the next block
             while ( j < k ) do
               If $C_j$ f $Q_i^j$
               // $i^{th}$ request be served
               then    $Q_i^{(j)} = Q_i^{(j)} + 1$;
               else  { j=j+1; continue }
             end while
      // No free port for the $i^{th}$ request
      end for
```

## 4. Scenario description for simulation parameters

We have considered the ports of the video on demand server dynamically divided into 20 to 25 partition blocks. Each partition block dynamically contains 10 to 8 numbers of ports to holds the client request. Clearly each block of partition is assigned to the different cluster of users population. The user assigned to a particular port means a link is assigned to that user for accessing the video server. We have considered any user from any cluster of population holds the port of the server. The user holds the link for at most 120 seconds to access any specific video in steady session. The user holds the link for at most 80 seconds for some interactive session. The overall simulation runs for 400 seconds.

## 5. Experimental Result and Discussion

Figure-2 and figure-3 represent blockage probability for the VoD demand Server. Fig-2 shows the blockage probability with respect to the arrival traffic rate. When the initial request came from the different cluster of population with rate 1, 1.5, 2, 2.5, 3, 3.5, 4, 4.5, 5 .For the first arrival of traffic request for the VoD movies it will forwarded to the first partition of the VoD – on-demand server. If all the ports of the $1^{st}$ partition were busy then the request will be forwarded to the next partition of the server.

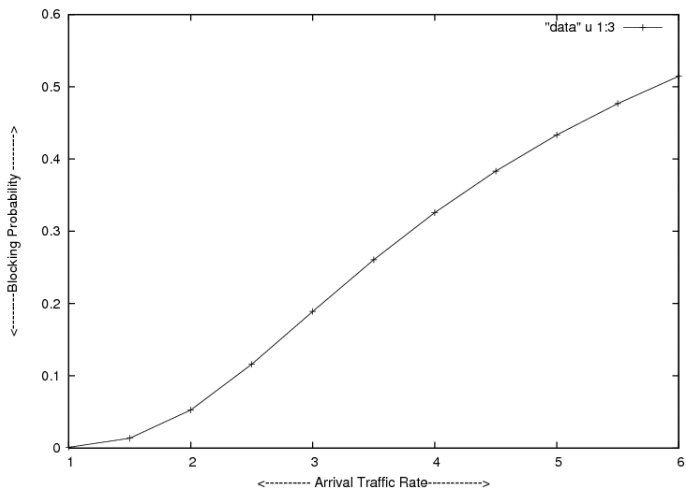

Figure -2

Figure- 3 clearly represents the blockage probability with respect to the traffic intensity inside the Network. We see that, when the initial traffic intensity is very low inside the VoD Network system the corresponding blockage probability was less. When the traffic intensity increases the number of free port becomes reduced and the blockage increased.

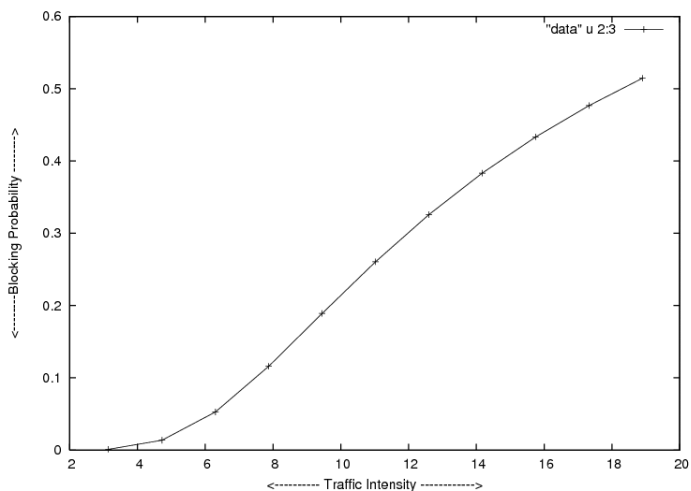

Figure -3

Figure-4 and figure -5 represents the VoD server performance with respect to the huge incoming traffic and the corresponding blockage probability. From fig-4 we see that when the incoming traffic rate is below 50 the performance of the VoD server was healthy but when the incoming rate is more than 200, the available free ports approaches to zero.

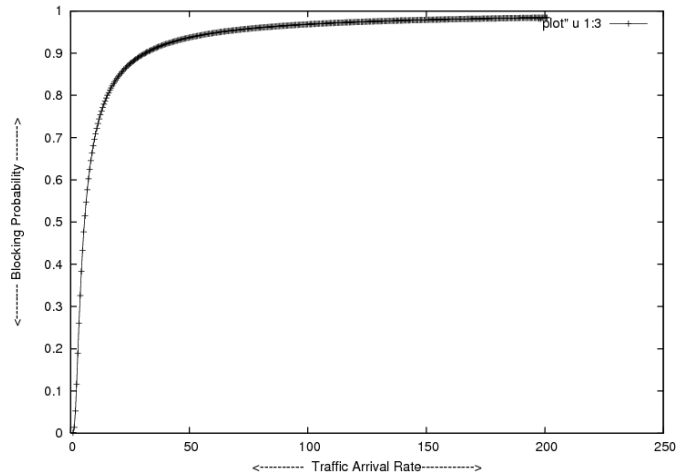

Figure -4

Figure-5 represents the VoD server performance with respect to the traffic load or the traffic intensity. Initially when the traffic load inside the VoD Network is less than 200 requests the VoD server worked well. But when the load inside the system increased, the blockage probability increased i.e. the available free port reduces. When the traffic load is more than 600 the blockage probability approaches to 1 i.e. there is no available free port to handle any more requests. With the increase of traffic intensity, packet loss increases.

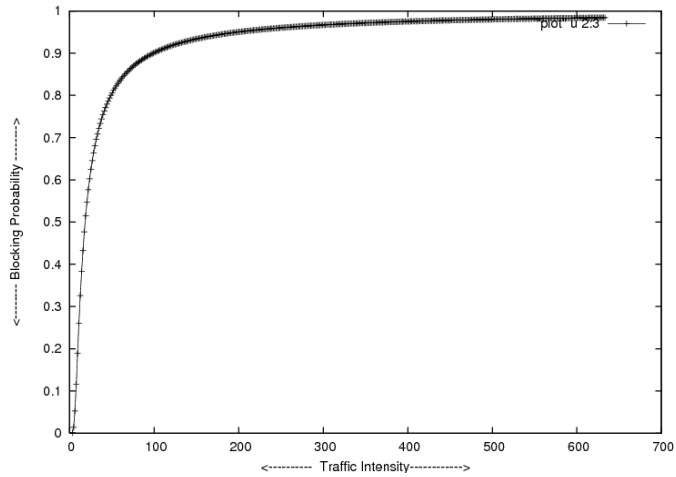

Figure -5

Figure 6 shown that the compared blockage probabilities of two situation. Line 1 represents the blockage probability with respect to the traffic load inside the VoD Server, where each partition of the server contains at least 10 ports i.e. $C_i > 10$. Here $i$ represents the $i^{th}$ partition of the server. Line 2 represent the blockage probability with respect to the traffic load in the VoD server. Line 2 corresponds to the partition of the VoD server that contain at least 8 ports i.e. $C_i > 8$. The graph clearly represents that when the number ports are less the blockage probability grows very rapidly.

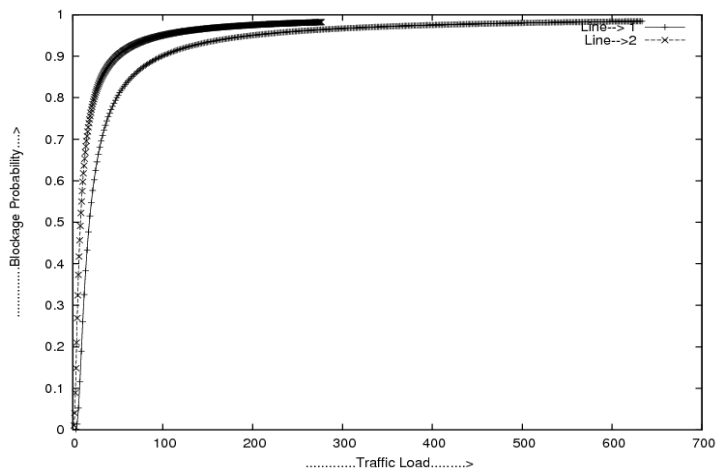

Figure -6

Figure-7 represents the traffic intensity with respect to the incoming traffic rate. Figure- 7 represents the traffic load in the VoD Network. The traffic load increases directly proportional to the arrival traffic rate to the VoD Server.

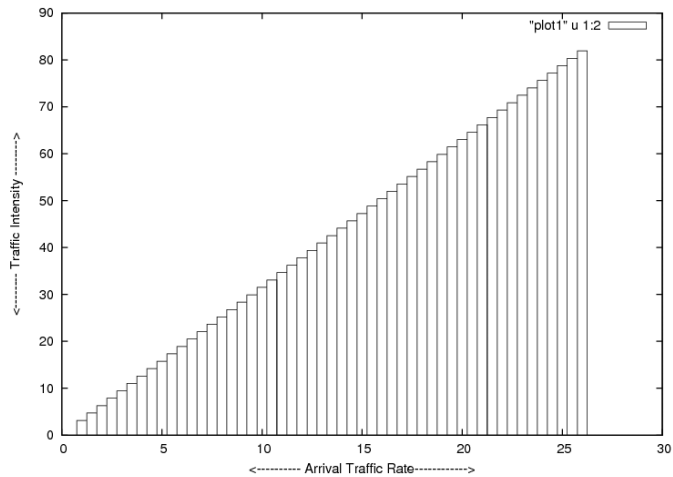

Figure -7

## 6. Concluding Remarks:

In this work we have considered a real scenario that appears in the video on demand server. In this work we have developed a novel algorithm for the packet forwarding to the free port in the VoD server. In this paper, we have analyzed the performance of the VoD system in the context of differential treatment of video requests. We have presented a methodology which can estimate the maximum load that the VoD server can handle for the request which comes from different class with different arrival rates. Through analytically and by simulation we showed that "partition based sharable on demand video server" admission control is a useful tool for manipulating performance for the request classes. The simulation result clearly reflects how the blockage probability affects the performance of the overall VoD server. For further work we plan to develop the procedure to implement the procedure into the Ad Hoc network. In that case the algorithm and the method of implementation has to be modified.

## 7. Acknowledgment:

The authors would like to thank Dr Sami for his suggestion.

[22]    K, Keeton and R. H. Katz. Evaluating video layout strategies for a high-performance storage server. Multimedia Systems, 3:43-52, 1995.

[23]    Y. Oyang, M. Lee, C. Wen, and C. Cheng. Design of multimedia storage systems for on-demand playback. In Proc. of Int'l Conf. on Data Engineering, pages 457- 465, Taipei, Taiwan, March 1995.

[24]    C. C. Aggarmal, J. L. Wolf, and P. S. Yu. A permutation-based pyramid broadcasting scheme for video on-demand systems. In Proc. of the IEEE Id? Conf. on Multimedia Systems '96, Hiroshima, Japan, June 1996.

[25]    S. Viswanathan and T. Imiehnski. Metropolitan area video-on-demand service using pyramid broadcasting. *Mulitmedia Systems,* 4(4):197-208, August 1996.

[26]    W. J. Bolosky, J. S. Barrera, R. P. Draves, R. P. Fitzgerald, G. A. Gibson, M. B. Jones, S. P. Levi, N. P. Myhrvold, and R. F. Rashid. The tiger video fileserver. In Proc. of the 6th Int'l Workshop on Network and Operating System Support for Digital Audio and Video, April 1996.

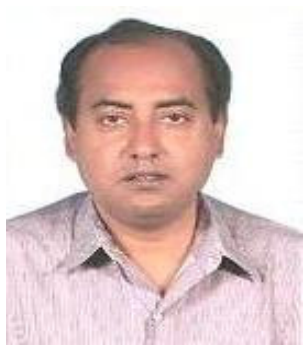

**Soumen Kanrar** received the M.Tech. degree in computer science from Indian Institute of Technology Kharagpur India in 2000. Advanced Computer Programming RCC Calcutta India 1998. and MS degree in Applied Mathematics from Jadavpur University India in 1996.
During 2001-2003 he worked in Techno India as faculty in computer-engineering department. During 2003-2005- he worked in Dr. B.C.Roy Engineering college India as faculty in the Computer Science and Engineering department. During 2005-2008 he worked in Durgapur Institute of Advanced Technology India as faculty in the Department of Computer Science and Engineering. Currently working as a Research Associate in the College of Computer & Information Sciences, King Saud University, Riyadh, Saudi Arabia

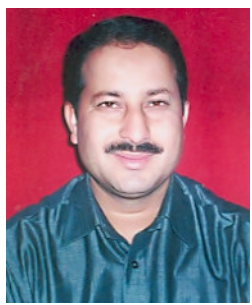

**Mohammad Siraj** received his M.E degree in Computer Technology and Applications from Delhi College of Engineering ,Delhi-India in 1997 and BE degree in Electronics and Communication Engg from Jamia Millia Islamia, New Delhi in 1995. He has worked as Scientist in Defence Research and Development Organisation (DRDO) Ministry Of Defence, India from 1995-2000. Currently he is working as a Lecturer in the College of Computer & Information Sciences, King Saud University, Riyadh, SA